\documentclass[11pt,a4paper]{article}

\usepackage[utf8]{inputenc} 
\usepackage[T1]{fontenc}    
\usepackage{microtype}      
\usepackage{amsmath,amssymb,amsfonts,amsthm} 
\usepackage{caption}
\newcounter{movie}
\newcommand{\movie}[1]{%
    \refstepcounter{movie}%
    \par\noindent\textbf{Movie S\arabic{movie}.} #1\par\vspace{0.75em}%
}
\usepackage[margin=1in]{geometry} 
\usepackage{graphicx}       
\usepackage{booktabs}       
\usepackage{hyperref}       
\usepackage{cleveref}

\hypersetup{
    colorlinks=true,
    linkcolor=blue,
    filecolor=magenta,      
    urlcolor=cyan,
    citecolor=blue,
}

\title{\textbf{Traveling Waves Enhance Transport during Cytoplasmic Streaming in \textit{Drosophila} Oogenesis}}

\author{
  Olenka Jain$^{a,b}$,
  David B. Stein$^{b}$,
  Reza Farhadifar$^{b}$,
  Elizabeth R. Gavis$^{c}$,\\
  Stanislav Y. Shvartsman$^{a,b,c}$,
  Michael J. Shelley$^{b,d,\dagger}$\\[0.5em]
  \small $^{a}$Lewis-Sigler Institute for Integrative Genomics,
  Princeton University, Princeton, NJ 08544, USA\\
  \small $^{b}$Center for Computational Biology,
  Flatiron Institute, New York, NY 10010, USA\\
  \small $^{c}$Department of Molecular Biology,
  Princeton University, Princeton, NJ 08544, USA\\
  \small $^{d}$Courant Institute, New York University,
  New York, NY 10012, USA\\[0.5em]
  \small $^{\dagger}$Corresponding author: mshelley@flatironinstitute.org
}
\date{\today}

\begin{document}

\maketitle
 
\begin{abstract}
Cytoplasmic streaming is believed crucial to large developing cells, such as oocytes, where flow provides a much faster route to transport and mixing of cellular components than molecular diffusion. Molecular motors carrying cargoes on cytoskeletal elements are known to drive streaming in a variety of biological systems and, in \textit{Drosophila} oocytes, cell-spanning vortical flows are tied to kinesin motors moving on dense beds of microtubules. Recent theories and simulations suggest such streaming self-organizes through interacting microtubules, motors, and flow. Whether these theoretically derived flows are adequate to the tasks of transport and mixing has been unclear. Here we report on new observations of coherent waves traveling persistently through microtubule beds during streaming, and investigate their impact on fluid transport and mixing. Leveraging recent advances in simulating the complex fluid-structure problems of cellular environments, we probe a biophysical model set in an oocytal geometry. Our simulations identify previously unknown waving states that show concordance with our experimental observation and the capacity for functional transport and mixing.
\end{abstract}

\vspace{0.5cm}

\section{Introduction}
Fluid dynamics has emerged as a powerful tool for understanding self-organization in large cells where cytoplasmic streaming is often needed for transport, mixing, and localization at the relevant timescales. In other words, cytoplasmic streaming seems to be one answer to the question ``how to organize a big cell?'' \cite{Goldstein2008,Drechsler2017}. Cell-scale fluid flows occur in large cells across the phylogenetic spectrum, from giant algal cells to zebrafish embryos to \textit{Drosophila} oocytes\cite{Okabe2008,Quinlan2016,Shimmen2007,Lu2018,Lu2022GoWithTheFlow,Pickard1974}. The particular purposes of the flow differ, whether they are important in the positioning of the spindle, the recirculation of metabolites, or the transport of maternal signaling factors \cite{Zimyanin2008OskarTransport,Sinsimer2013,kugler2009localization,Frohnhofer1986bicoid,Okabe2008}, but the ingredients of flow generation appear to be conserved. In general, cytoplasmic streaming appears to involve cytoskeletal filaments (such as microtubules or actin) and their associated molecular motors (such as dyneins, kinesins, and myosins) \cite{Goldstein2015,Shamipour2019BulkActin,Lu2016,TheurkaufEtAl1992}. This conserved biophysical structure has allowed fluid-fiber interaction models to be especially powerful tools for understanding both the generation and function of cytoplasmic streaming \cite{Chakrabarti2022,Htet2025Analytical,PhysRevFluids.9.120501,Tornberg2004,Stein2021}.

\begin{figure*}[t!]
\centering
\includegraphics[scale=0.62]{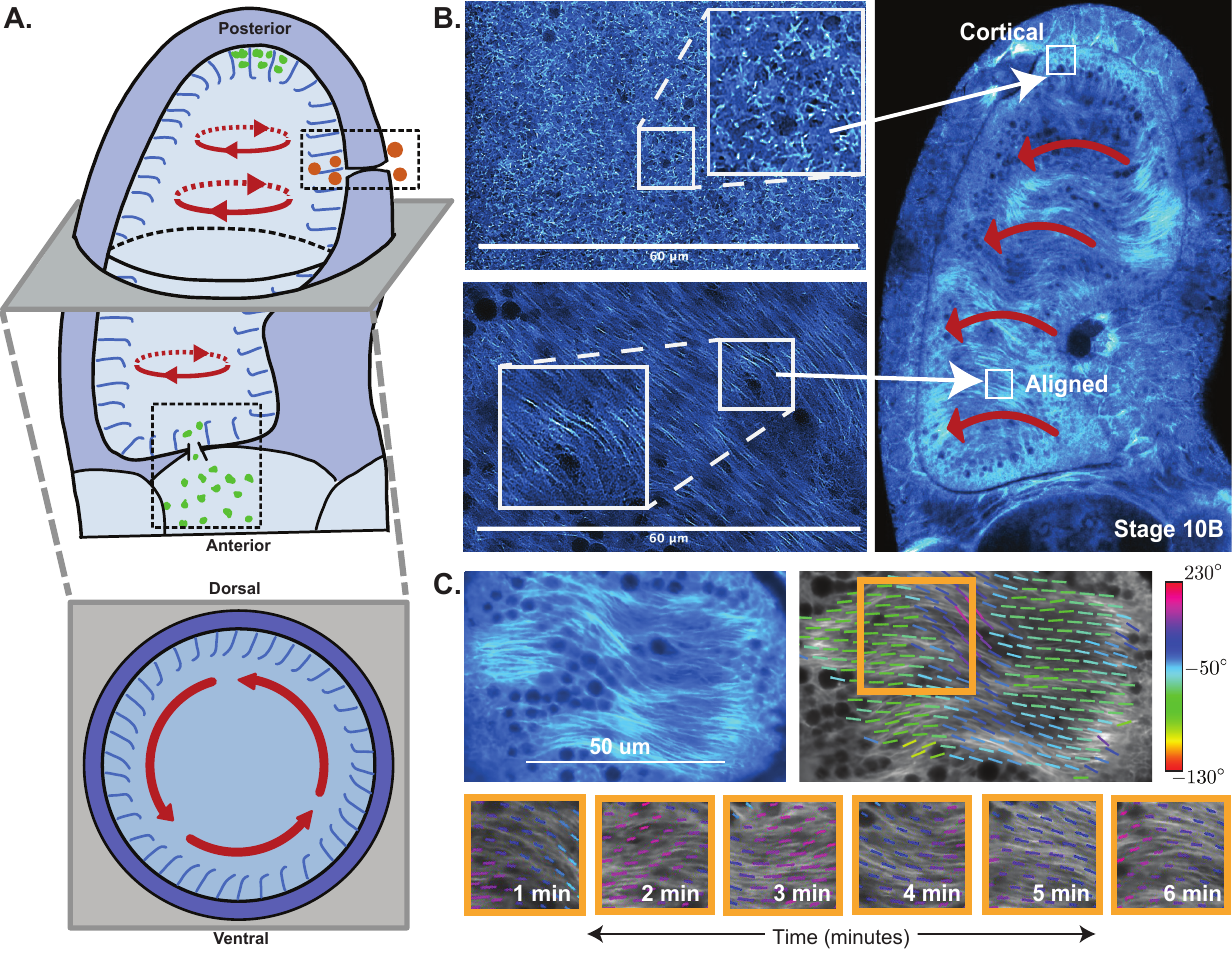}
\caption{A. Cartoon representing cortically anchored microtubules (blue fibers) in the late-stage oocyte (stage 10B). Vortical flow often rotates around the long axis of the cell. Anteriorly located nurse cells provide important factors, such as the maternal RNA \textit{nanos} (represented by green particles), that must be localized at the posterior pole of the oocyte \cite{Wang1991Nanos}. Yolk (represented by brown particles) endocytoses through the follicular epithelium surrounding the oocyte. B. (right) A max projection of $10 \ \mu \mathrm{m}$ thick cortical slices of a stage 10B oocyte during cytoplasmic streaming where the microtubules were visualized using the Jupiter-GFP \textit{Drosophila} line \cite{Karpova2006}. (left) Super-resolution imaging \cite{Prakash2022} of another oocyte shows the general structure of the microtubule bed, which is (top) largely disorganized up to $5 \ \mu \mathrm{m}$ from the cortex and becomes (bottom) highly aligned with the flow $5-10 \ \mu \mathrm{m}$ from the cortex. 
C. The Gabor filter was used to detect the dominant local microtubule direction on a set grid inside the oocyte (color-coded line segments). The temporal change in microtubule orientation can be visualized by tracking the orientations within a particular section over time (orange insets).}
\label{fig:figure1}
\end{figure*}

The volume of the \textit{Drosophila} oocyte grows approximately $100,000$-fold during the course of oogenesis, from a few $\mu \mathrm{m}$ diameter sphere to an approximately prolate ellipsoid shape, with a long axis reaching $500\ \mu \mathrm{m}$ \cite{OFarrell2015GrowingEmbryo,Miles2011ArtificialSelectionEggSize,deCuevas2015_drosophila_oogenesis}. During its period of growth, the oocyte must sort and localize important maternal signaling molecules from bordering nurse cells. Streaming is necessary for the timely transport of maternal cues from the anteriorly located nurse cells to the posterior pole \cite{Becalska2009LightingUpmRNA,Jaramillo2008Gurken,Gregor2007BicoidGradient,Driever1988Gradient}. When the oocyte reaches approximately $150\ \mu \mathrm{m}$ in length at stage 10B \cite{Jia2016Automatic}, cell-scale cytoplasmic streaming occurs, characterized by a coherent rotational motion of the cytoplasm \cite{Drechsler2020,Gutzeit1982,Forrest2003NanosLocalization,Quinlan2016}. Additionally, streaming appears to uniformly disperse, within the cytoplasm, yolk granules that are endocytosed at the cortex during the process of vitellogenesis (stages 8-10) \cite{Schonbaum2000_yolkless,Yu2024EndolysosomalTrafficking, Bownes1993_regulation, Schonbaum2000_yolkless, ramos2022open,Forrest2003NanosLocalization, Kilwein2023DrosophilaEmbryos}. Without streaming, the dumping process, whereby nurse cells expel their entire contents into the oocyte, pushes yolk granules back to the cell cortex where they remain segregated. This segregation presumably causes future problems in the embryo, for example, preventing the syncytial nuclei from reaching the cortex [\cref{fig:figure1}, A]. 

Experimental and simulation-based evidence has led to a theoretical model, referred to here as the Twister model, whereby cortically anchored microtubules, forced by molecular motors, drive cytoplasmic streaming \cite{Stein2021,PhysRevFluids.9.120501,Dutta2024SelfOrganizedTwisters,varchanis2026stabilized,Ganguly2012,Parton2011PAR1,Gutzeit1982,Gutzeit1986,Serbus2005,Dahlgaard2007,Monteith2016}. Therefore, we sought to characterize the dynamics of microtubules during streaming using live imaging. While previous studies have confirmed that microtubules are aligned with the local flow, they have not characterized microtubule dynamics during streaming or on longer time scales. Here we report that during cytoplasmic streaming, deformation waves travel through the microtubule bed. We show that similar waves appear in a specific parametric regime of the Twister model \cite{Dutta2024SelfOrganizedTwisters, Jain2025GeometricEffects, Stein2021,PhysRevFluids.9.120501}. Our capacity to probe this complex model relies on new advances in numerical methods \cite{stein2024computational}. Finally, based on these simulational studies and experimental perturbations of the flows, we argue that the waving dynamics can provide the material transport of mRNA transverse to the large-scale vortical flow important for axis specification, while secondary flows inwards from the cortex could support re-suspension, after dumping, of yolk granules which serve as the source of protein for the future embryo \cite{Swevers2005Vitellogenesis,Lin1993GermlineStemCell,Raikhel1992Accumulation,Kilwein2023DrosophilaEmbryos}.

\section*{Jupiter-GFP Illuminates Cortical Microtubules}

Previous reconstructions of microtubule structure during late stage cytoplasmic streaming were based on short snapshots or fixed oocytes \cite{Theurkauf1993,Quinlan2016,Dutta2024SelfOrganizedTwisters}. Microtubules were shown to locally align with the direction of streaming, yet their long term dynamics were not characterized. The Twister model explained these experiments through an attracting nearly-steady streaming state whose bed of microtubules are bent and wrapped around the vortical axis of the cell [\cref{fig:figure1}, A] \cite{Dutta2024SelfOrganizedTwisters,Stein2021,Jain2025GeometricEffects}.

To investigate the live microtubule dynamics we used confocal microscopy and a standard Jupiter-GFP \textit{Drosophila} line. Jupiter is a microtubule associated protein, often used to visualize microtubules as shown in [\cref{fig:figure1}, B]. It is important to note that live imaging using a confocal microscope allows for a maximum XY-resolution of ${\sim}260 \ \mathrm{nm}$, which is over ten times wider than a single microtubule. Therefore, these images are useful for characterizing large scale dynamics and overall microtubule alignment, not individual microtubule dynamics. Lastly, we focused on the thin ordered region of microtubules where microtubule alignment and signal is strongest. Microtubules visibly emanate from the cortex, and a region of microtubules with long range order exists ${\sim}5-10 \ \mu \mathrm{m}$ from the cortex. It is this aligned region that is believed to generate cytoplasmic streaming. Using super-resolution microscopy and microtubule segmentation, we characterized the thickness of this aligned region and found it to be only a few $\mu \mathrm{m}$s thick. Therefore, for the live imaging of waving microtubules, we used a maximum projection of a ${\sim}10 \ \mu \mathrm{m}$ thick slice which captures the aligned microtubule region [\cref{fig:figure1}, C]. 

To characterize the microtubule alignment we used a Gabor filter. The Gabor filter, composed of a sinusoid modulated by a Gaussian, is a common image processing tool for detecting edges \cite{Mehrotra1992Gabor}. Specifically, we applied a set of Gabor filters to a grid of points on each image as previously done in \cite{Dutta2024SelfOrganizedTwisters,Jain2025GeometricEffects}. The filter orientation with the highest resultant magnitude gave the local microtubule orientation at each gridpoint. This approach allowed us to extract an orientation field of the microtubule beds. Examples of such extracted fields are shown in [\cref{fig:figure1}, C]. Unlike the previous predictions of the Twister model, we found that microtubules were not stably aligned but instead showed traveling wave patterns. We found these traveling waves in every oocyte imaged. Thus, the live imaging and image analysis revealed time dependent orientation fields not previously observed in the Twister model. 
\begin{figure*}[t!]
\centering
\includegraphics[scale=0.7]{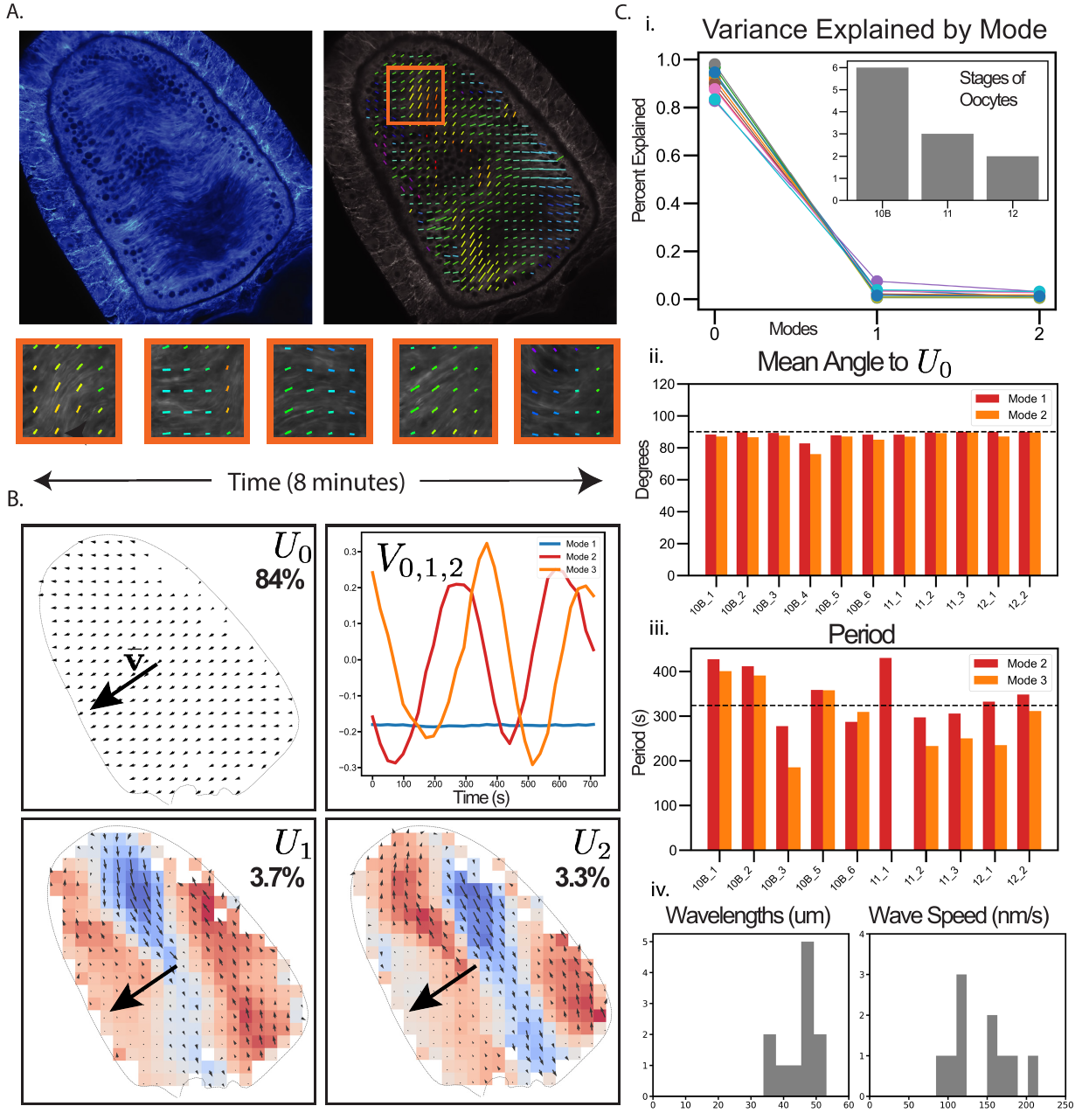}
\caption{A. Representative live confocal microtubule imaging using the Jupiter-GFP \textit{Drosophila} line and the resulting orientation field found using the Gabor filter. Orange insets show the orientation field at a fixed location over time. B. SVD decomposition for the vector field from A. The first spatial mode of the SVD gives the direction of stable streaming (up to a sign) depicted by the bolded vector, $\bar{\mathbf v}$. The second and third spatial modes have a characteristic structure in which the vectors form elongated patches along the long axis of the cell (shown by the blue and red coloring). The vectors of spatial modes 2 and 3 are mostly perpendicular to $\bar{\mathbf v}$. The time dependent amplitude of mode 1 is constant while the amplitudes of modes 2 and 3 are characterized by an out-of-phase, equal-period, wave. C. Statistics from 11 videos across the 3 stages of oocytal development during which cytoplasmic streaming occurs.  Oocytes whose imaging times were shorter than a period were not included. C.i The first mode consistently explains most of the variance in the spatiotemporal field while modes 2 and 3 contribute around $3-5\%$. The next singular value typically explained less than 1\%. C.ii The mean angle between mode 1 and modes 2 and 3 is consistently perpendicular across oocytes. C.iii-iv The period (${\sim}6$min), wavelengths (${\sim}40\mu$m), and wave speeds (${\sim}150$nm/s) are largely consistent across oocytes.}
\label{fig:figure2}
\end{figure*}

\section*{Identification of the Traveling Wave Regime}
As a first step in analyzing these spatiotemporal orientation fields we used the singular value decomposition (SVD) to decompose the field into a set of orthogonal spatial modes, {$U_i$}, modulated by time dependent amplitudes, {$V_i$}. The importance of each mode is captured by the singular values, $\sigma_i^2$, giving the variance explained by that mode \cite{Dutta2020SurfaceFlow,Embree2019SVD}.

An example is shown in [\cref{fig:figure2}, A-B]. Here, the first spatial mode, $U_0$, explains $84\%$ of the total variance and is governed by a nearly constant amplitude. The next two spatial modes are approximately equal in their importance ($\sigma_1^2\,{\approx}\,4\%,\ \sigma_2^2\,{\approx}\,3\%$) and capture the oscillating time dependent modes. These two modes are out of phase by $\pi/2$. The spatial modes, $U_1$ and $U_2$, have a characteristic structure. Nearly all vectors in these modes are perpendicular to the mean vector from $U_0$, $\bar{\mathbf v}$, so that each vector's orientation is either $\bar{\mathbf v} \ + \pi /2$ or $\bar{\mathbf v} - \pi /2$. Lastly, there is a segregated spatial structure with patches of $\bar{\mathbf v} + \pi /2$ pointing vectors and $\bar{\mathbf v} \ - \pi /2$ pointing vectors. 

Such a modal structure can be generated by a low amplitude traveling wave. For example, a 1D traveling wave with sufficiently low amplitude (or long wavelength) of the form $A\sin(kx-\omega t)$ will generate a dominant mode of vectors in the direction of travel followed by two equal amplitude modes perpendicular to the direction of travel. The spatial structure of $U_1$ and $U_2$ reveals that the wavelength and the period of their time dependent amplitudes is the period of the underlying wave. Lastly, the time dependent modes $V_1$ and $V_2$ are $\pi/2$ out of phase with one another. Further details about the results of SVD on vector fields generated by standing and traveling waves are described in the S.I.

In other words, the first mode of SVD, $U_0$, is consistent with microtubules stably bent in the direction of fluid flow. This dominant mode matches the dynamics of microtubules from previous numerical simulations in the stable streaming regime of the Twister model \cite{Stein2021,Dutta2024SelfOrganizedTwisters,Jain2025GeometricEffects}. These secondary modes suggest that the microtubules are not just stably bent and instead there is a low amplitude wave that travels around the cortical microtubule bed in the direction of flow.

The decomposition shown in [\cref{fig:figure2}, B] was not an isolated example, and the structure of the SVD modes was consistent across cells exhibiting streaming and even within the same cells at different time points in their development. The streaming direction was predominantly axisymmetric (around the long axis of the cell) as shown before in \cite{Jain2025GeometricEffects}. Statistics extracted from these modal decompositions for 11 oocytes across various stages of axisymmetric streaming are shown in [\cref{fig:figure2}, C]; these are largely independent of oocyte and stage. In some oocytes, the traveling wave was coherent along the length of the cell as shown in [\cref{fig:figure2}, C], while in others the traveling wave was de-cohered along the length of the cell, particularly in cases where the microtubule bed appeared patchy. In conclusion, the dynamics of the microtubule bed are low dimensional and can be explained by a dominant streaming regime modulated by a traveling wave of deformation.

\begin{figure*}[t!]
\centering
\includegraphics[scale=0.3]{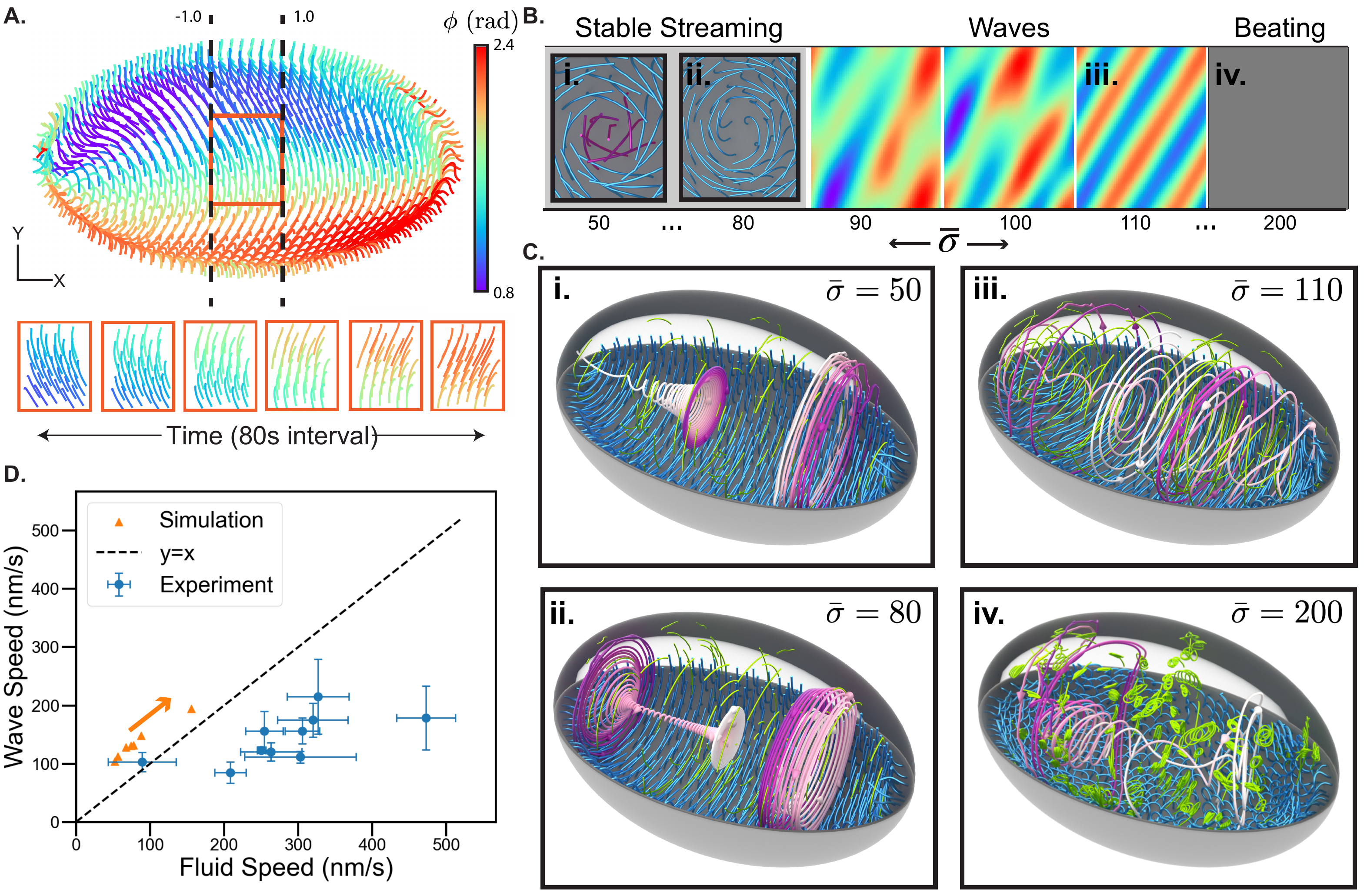}
\caption{A. Snapshot of a representative simulation in the traveling wave regime looking down at all microtubules below the mid-plane of the ellipsoid. Each microtubule is colored by the angle, $\phi$, it makes with the $x$-axis (the long axis of the ellipsoid). A series of snapshots for the orange boxed region illustrates the waving dynamics. B. The different behaviors observed as $\bar{\sigma}$ is varied over fixed $\bar{\rho}=15$. (i, ii) Stable streaming flows appear when $\bar{\sigma}\lesssim90$. (i) At low $\bar{\sigma}$, microtubules at the defect are nearly straight (purple). (ii) For larger $\bar{\sigma}$, the defect structure changes, with microtubules near to the defect splayed outwards. (iii) traveling waves appear for $90\lesssim\bar{\sigma}\lesssim110$. We characterize the traveling wave regime by considering the middle belt of microtubules (shown in the dashed black lines of A.). The traveling wave can be seen clearly with a kymograph that plots $\phi$ as a function of the circular cross-section of the belt, parameterized by $\theta$. (iv) For sufficiently high $\bar{\sigma}\gtrsim110$, the waves break down and yield to locally coordinated beating. C. Microtubule orientations (blue), streamlines (pink to purple in the direction of flow), and particle paths (light green) are shown for each regime below. The change in defect character from B.i to B.ii coincides with a reversal of the streamline direction between panels i and ii. Time-dependence of the flows leads to the separation of streamlines and particle paths in panel (iii) and, more dramatically, to panel (iv). D. In both experiment and simulation, fluid flow is on the order of the wave speed. In the simulations, fluid speed is a function of increasing $\bar{\sigma}$ (represented by the orange arrow).}
\label{fig:figure3}
\end{figure*}

\section*{The Twister Model of Cytoplasmic Streaming}

We hypothesized that these spatiotemporal patterns arise in a specific parameter regime of the Twister model, which poses the system as a geometrically complex fluid-structure interaction problem \cite{Stein2021,Dutta2024SelfOrganizedTwisters}. In the Twister model, microtubules are attached to the cortex of the oocyte and experience compressive load from plus-end directed kinesin-1 motors \cite{Lu2016,delCastillo2015KinesinDyneinDrosophila}. These motors carry cargo, which entrain the surrounding cytoplasm, towards the microtubules' free ends. The compressive load is represented by a constant line density of force, with magnitude $\sigma$, along each of the $N\gg1$ microtubules. The position of the $i^\textrm{th}$ microtubule is given by $\mathbf{X}^{i}(s,t)$, where $s$ is arclength from its anchored and clamped base, and $t$ is time. The microtubule moves in a background flow $\bar{\mathbf{u}}_i$, and its dynamics are given by:
\begin{align}
\eta (\mathbf{X}^i_{t}-\bar{\mathbf{u}}_{i}) &= (\mathbb{I}+ \mathbf{X}_{s}^{i} \mathbf{X}_{s}^{i}) \cdot (\mathbf{f}^i - \sigma \mathbf{X}_{s}^{i}), 
\\
\mathbf{f}^i &= -E \mathbf{X}_{ssss}^{i} + (T^{i} \mathbf{X}_{s}^{i})_{s}.
\end{align}
Each microtubule is taken to have the same time-independent length, $L$. The background flow $\bar{\mathbf{u}}_i$ captures the nonlocal contributions to the velocity field from every other microtubule (and the motors dragging cargo along it) and $\mathbb{I}$ is the identity tensor. The force per unit length, $\mathbf{f}^i$, includes both bending and tensile components; here $E$ a bending rigidity and $T^{i}$ the tension, which acts as a Lagrange multiplier to enforce the local inextensiblity of microtubules (i.e. $\mathbf{X}_s^i\cdot\mathbf{X}_s^i$=1), with $s$ subscripts denoting arc-length derivatives. The drag coefficient is given by $\eta=8 \pi \mu/c$, where $\mu$ is the viscosity and $c=|\log e (r/L)^2|\gg1$ is a coefficient characterizing the slenderness of the microtubule, with $r$ and $L$ its radius and length, respectively. Each microtubule is clamped normally to the cortex and the free end is both force and torque-free. 

The fluid flow is governed by the forced incompressible Stokes equation, 
\begin{align}
\nabla q - \mu \Delta \mathbf{u}=  \sum\limits_{i=1}^N \int\limits_{0}^{L} \mathrm{d}s \  \mathbf{f}^{i}(\textit{s})\delta(\mathbf{x}-\mathbf{X}^{i}(s));\ \nabla \cdot \mathbf{u}= 0
\end{align}
where $q$ is the pressure, and subject to a no-slip boundary condition on the fluid velocity $\mathbf{u}$ at the cortex.

This model comprises a complex fluid-structure interaction problem, with each microtubules' conformation, $\mathbf{X}^{i}(s,t)$ described by a stiff partial differential equation and coupled together by the ooplasmic flow that they collectively force [eqs. 1-3]. When there are hundreds to thousands of microtubules, evolving this system is a complex task requiring specialized methods and software, \textit{SkellySim}, which are described in the Methods. 

The behavior of the system is largely governed by two dimensionless groups:
\begin{equation}
\bar{\rho} = \frac{8 \pi N L^2}{cS}, \quad
\bar{\sigma}  = \frac{\sigma L^3}{E},
\end{equation}
where $\bar{\rho}$ is the non-dimensional microtubule density, $\bar{\sigma}$ is the non-dimensional motor force per-unit-length on the microtubules, and $S$ is the cortical surface area. With large enough motor forcing {$\bar{\sigma}$}, isolated microtubules buckle and bend, resulting in time-dependent beating or rotations \cite{DeCanioLaugaGoldstein2017,clarke2024bifurcations}. However, the motion of the microtubules is coupled through hydrodynamic interactions in the cytoplasm, and when the density of the microtubules per unit surface area ($\bar{\rho}$) is high enough, the microtubules bend collectively to form an aligned and nearly stationary streaming state \cite{Stein2019,Dutta2024SelfOrganizedTwisters,re:Ohm:phd}. 

\begin{figure*}[t!]
\centering
\includegraphics[scale=0.23]{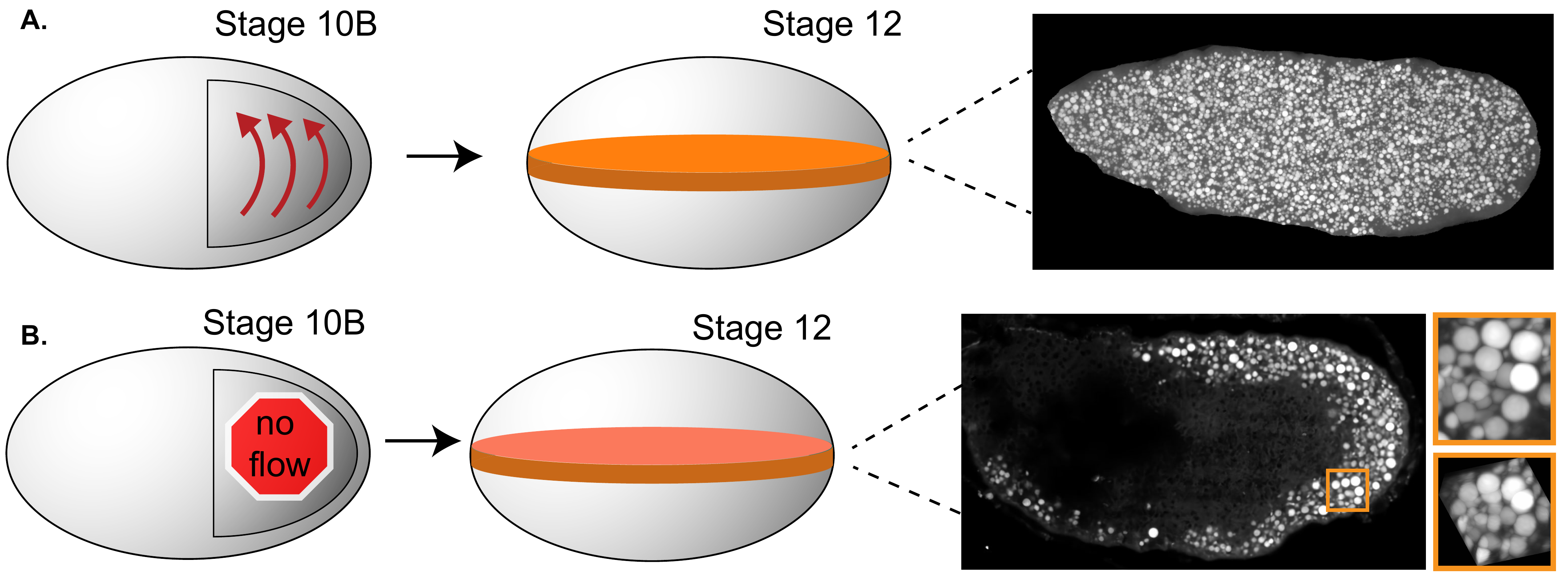}
\caption{Two conditions were used to investigate the effects of cytoplasmic streaming on spatial yolk distribution. A. Stage 10B oocytes were allowed to stream normally, dissected at stage 12-14, and cryosectioned to visualize the middle $7 \mu \textrm{m}$ slice, showing yolk suspended with apparent uniformity. B. Stage 10B oocytes were treated with colchicine which inhibits streaming, cultured until stage 12-14, and cryosectioned to visualize the middle $7 \mu \textrm{m}$ slice. The yolk granules are now sharply segregated to the posterior cortical region, presumably a consequence of flows from nurse cell dumping in stage 11. The two right-hand boxes illustrate the tight packing of the segregated yolk granules from different angles of the 3D volume.}
\label{fig:figure4a}
\end{figure*}

Linear stability analysis and nonlinear simulations of the Twister model using \textit{SkellySim1.0} three regimes as a function of these governing parameters: (1) \textbf{The stationary stable regime} where microtubules are unbent and the fluid is quiescent, (2) \textbf{The stable streaming regime} where rotational cell-spanning flows form around the long axis of the cell. In the streaming regime, microtubules are bent and aligned with the direction of the flow (except at the poles of the cells, where there are two +1 order orientational defects in the projected surface vector field of the microtubules). The presence of the defects and the closed geometry results in an additional bitoiroidal component of the flow, which re-circulates particles within each hemisphere and whose magnitude is much less than that of the rotational flow \cite{Dutta2024SelfOrganizedTwisters}. And, (3) \textbf{The beating regime} where microtubules beat in a globally uncoordinated manner and the mean flow velocity is small.  We reasoned that if a traveling wave regime were to be captured by the Twister model, it would appear around the boundary of streaming and beating.

\section*{A Traveling Wave Regime Emerges from the Model}

To explore this system here, we used a yet more advanced software suite, \textit{SkellySim2.0}, to simulate the hydrodynamics of compressively loaded microtubule beds confined within ellipsoidal cellular geometries, chosen to match the shape of late-stage oocytes. This new software features substantial methodological improvements (see Methods) which allowed us to explore regimes of the system inaccessible to \textit{SkellySim1.0}, the software used in \cite{Dutta2024SelfOrganizedTwisters,Jain2025GeometricEffects,skellysim}

Our simulations revealed a new regime corresponding to traveling waves in the microtubule bed. The waves arise when the aligned microtubules of the streaming regime begin to ``sway'' transversely to the direction of streaming. While the microtubules at the two defects individually beat (like microtubules in the beating regime), the defects remain near the poles of the ellipsoid. The waves extend along the long axis of the ellipsoid and travel in the direction of the underlying twister flow, with an amplitude that grows with the characteristic motor force, $\bar{\sigma}$. The underlying flow is still largely axisymmetric vortical flow, now just modulated by the transverse movements of the microtubules. The coherent traveling wave is lost as the system transitions to the beating regime, where the defects become dislodged from the poles and even proliferate, showing line defects among others.

An example of a simulation in the traveling wave regime is shown in [\cref{fig:figure3}, A]. To characterize the traveling waves, we focused on microtubules in the equatorial region of the ellipsoid, measuring the angle $\phi$ they make with the long axis of the ellipsoid [\cref{fig:figure3}, A]. In stable axisymmetric streaming, these microtubules are perpendicular, with $\phi=\pi/2$, to the long axis \cite{Jain2025GeometricEffects}. In the traveling wave regime, the microtubules deviate from $\phi=\pi/2$. To make connection with our experimental observations, any simulation in which the central region displayed deviations greater than $5^\circ$ was characterized as ``waving.'' Because the wave travels around the circular cross-section of the prolate ellipsoid, we plotted $\phi$ as a function of $\theta$, the azimuthal angle on the circular cross-section. We used a kymograph of $\theta$ as a function of simulation time to visualize the traveling wave as seen in [\cref{fig:figure3}, B] (See SI for more details). The simulation time is made non-dimensional by the single microtubule relaxation time, $\tau_r=\eta L^4/(cS)$, which is identical for all simulations. The diagonal lines of red and blue correspond to peaks and troughs traveling around the ellipsoid. We calculated wave-speed from the movement speeds of these extrema, wavelength from the distance between extrema, and the amplitude from the range of $\phi$. Interestingly, in both experiment and simulations, wave speeds are on the order of cytoplasmic streaming speeds, though the experimental were consistently higher [\cref{fig:figure3}, D] \cite{Shelley1992CoherentStructures}. Lastly, by running a parameter sweep over $\bar{\rho}$ and $\bar{\sigma}$, we found that the traveling wave exists between the streaming and beating regimes. We confirmed that this region is robust to the geometry of the cell by performing the same numerical simulations in various aspect-ratio ellipsoids, including a sphere (see SI). 

\begin{figure*}[t!]
\centering
\includegraphics[scale=0.12]{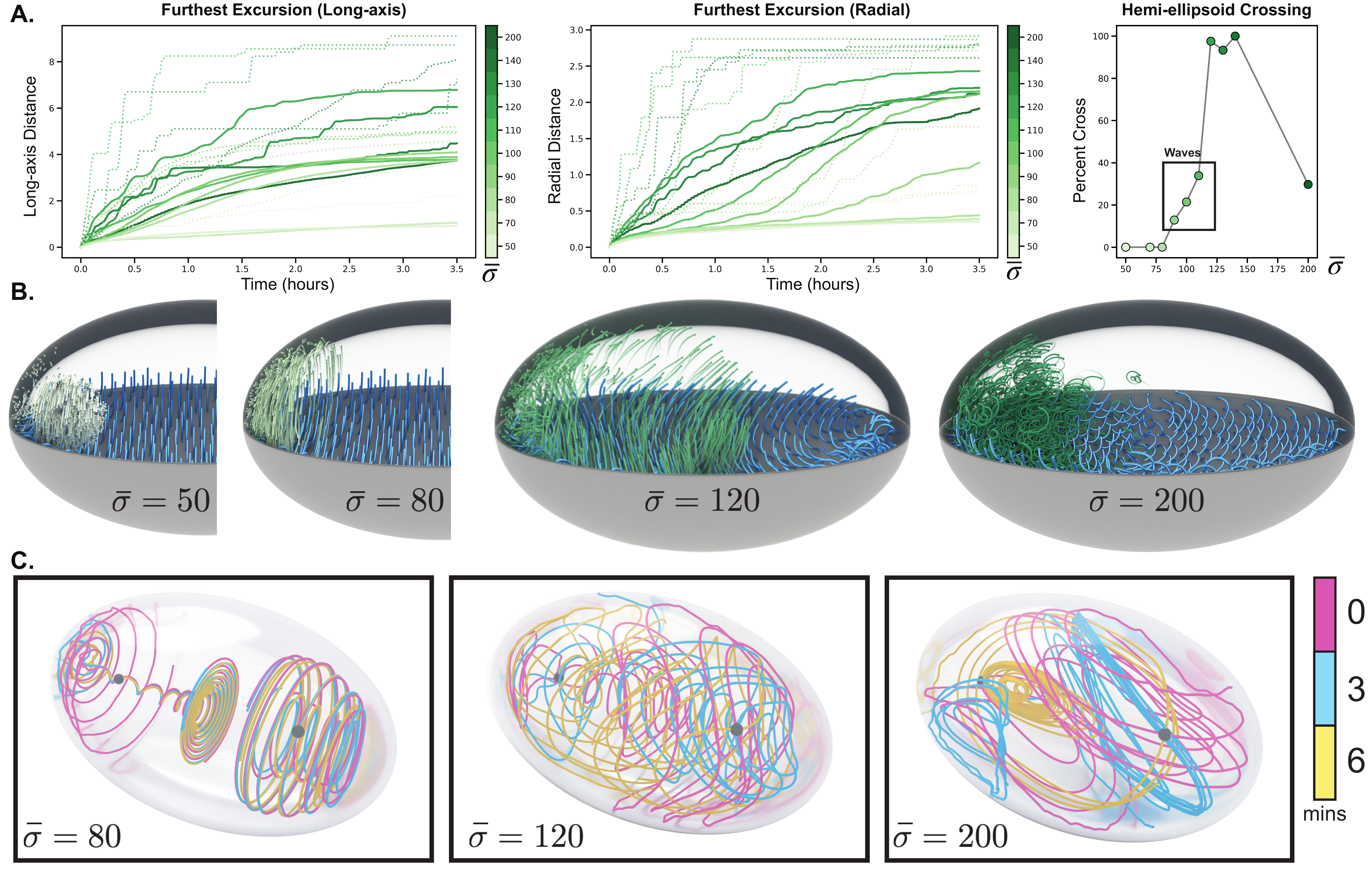}
\caption{A. The furthest excursion along the $x$-axis of a tracer particle initialized in the posterior tenth of the ellipsoid is plotted as a function of time. The mean furthest excursion is represented by solid lines and the maximum by dotted lines. The traveling wave regime occurs for $90\lesssim\bar{\sigma}\lesssim110$ and shows significantly increased long-axis transport than does stable streaming, where particles never leave their initial hemi-ellipsoid. Likewise, the speed of radial transport of particles initialized in a belt with initial radius larger than $5R/6$ from the centerline is greater in waving than stable flows. Finally, we choose to measure the percent of posteriorly initialized particles that cross over by a small threshold (0.2) into the opposite hemi-ellipsoid and find a peaked function of $\bar{\sigma}$ with the traveling wave regime under the left half of the peak. B. Path lines of posteriorly initialized particles after ${\sim}50$ minutes of flow. Low $\bar{\sigma}$ stable flows show poor transport, even as the direction of bitoroidal streaming is flipped at $\bar{\sigma}=80$. Likewise, very high $\bar{\sigma}$ flows around $\bar{\sigma}=200$ show reduced transport capacity as particles get stuck in local orbits. C. Streamlines at three times, seeded from the same points (black dots). At $\bar{\sigma}=80$, flows are nearly independent of time, with a separatrix at the midplane of the ellipse. In the traveling wave regime ($\bar{\sigma}=120)$, each set of streamlines has a similar appearance, but the there is no such constant separatrix, enabling particles to cross into otherwise inaccessible regions of the cell. In the beating regime ($\bar{\sigma}=200)$, the time dependence is more profound, with both the flow character and direction changing rapidly.}
\label{fig:figure4b}
\end{figure*}

Not only did the simulations reveal traveling wave dynamics, they additionally shed insight on the transition to traveling waves from stable streaming. Even in the stable streaming regime prior to the onset of waves, the microtubules at the defect change their dynamics as a function of $\bar{\sigma}$. In stereotypic stable streaming [\cref{fig:figure3}, B.i], microtubules at the defect point inwards, as reported previously \cite{Dutta2024SelfOrganizedTwisters}. As $\bar{\sigma}$ increases and the system approaches the traveling wave regime, microtubules at the defect instead splay outwards [\cref{fig:figure3}, B.ii]. This change in defect type corresponds to a flip in the direction of the bitoiroidal component, with cytoplasm at the pole now recirculating along the edge of the cell and coming back along the centerline. This change in defect structure with the flip in bitoiroidal flow direction is seen in [\cref{fig:figure3}, C.i-ii]. It is plausible that this change in direction of recirculation could be especially useful for the biological system which generally needs to localize maternal components to particular regions of the cortex.
 
\section*{Enhanced material transport in the traveling wave regime}

We hypothesized that the biological system lives in the traveling wave regime because the flow has different properties than the vortical flow of the streaming regime. Specifically, we considered the transport properties of streaming flows versus waving flows based on the known biology of the \textit{Drosophila} oocyte. Prior to the onset of cytoplasmic streaming, yolk vitellogenesis begins in the oocyte (stage 8) \cite{Schonbaum2000_yolkless, Bownes1982_vitellogenesis,Row2021OocyteLipidIntake}. The yolk proteins are brought into the oocyte through endocytotic machinery at the cortex of the oocyte and repackaged into yolk granules. By the end of oogenesis, yolk granules are uniformly distributed throughout the \textit{Drosophila} oocyte. Prior experiments have suggested that in the absence of streaming, the yolk granules remains segregated to the cortex \cite{Forrest2003NanosLocalization}. Thus cytoplasmic streaming is potentially important not only for transport of maternal RNA, but also for ensuring a uniform distribution of yolk granules throughout the early embryo \cite{Kilwein2023DrosophilaEmbryos,Tanaka2021_yolk_oskar, kugler2009localization}.  
 
We experimentally visualized the effects of cytoplasmic streaming on yolk granule suspension. Previous experiments which observed segregated yolk granules upon streaming inhibition were not able to image deep into the oocyte due to its optical density. Therefore, to quantify the distribution of yolk throughout its whole volume, we imaged $7 \mu \mathrm{m}$ thick sequential slices using a cryostat and confocal imaging. We incubated oocytes at stage 10B with microtubule de-polymerizing drugs to prevent streaming, and compared the resulting distributions with control oocytes that streamed normally \cite{Forrest2003NanosLocalization}. The differences were striking. Earlier stage 10B oocytes showed uniform suspension of yolk, whether streaming was inhibited or not. However, during and after nurse cell dumping in stage 11, inhibited oocytes showed a sharp segregation of yolk to the posterior cortex, while normally streaming oocytes showed a uniform distribution [\cref{fig:figure4a}]. We hypothesize that dumping pushes the yolk aside to the cortex, and without fast streaming, it cannot re-suspend [\cref{fig:figure4a}, B].

The observation that cytoplasmic streaming apparently redistributes yolk within the oocyte led us to examine the transport properties of streaming versus waving flows from simulations. We found that the capacity for both radial and axial transport capacity of waving flows was far greater than that of stable streaming flows. There are a few features of the stable streaming flows that make them poor transporters. One is the existence of a ``separatrix'' in the flow, a plane across which particles cannot cross. In the case of a prolate ellipsoid, that plane occurs at the circular cross-section at the midpoint of the long axis, and limits particles to their hemi-ellipsoid of origin \cite{chakrabarti2024cytoplasmic,Jain2025GeometricEffects,Dutta2024SelfOrganizedTwisters}. Because the streamlines are largely time-independent in the stable streaming regime, the separatrix does not move. This is in contrast to the waving flows, whose streamlines are time dependent [\cref{fig:figure4b}, C]. 

To consider both the long-axis and radial transport of particles in the various flow regimes we seeded tracer particles in the simulation at one end of the ellipsoid (at up to one tenth of its length) and around a ``radial belt'' (in the center fifth of the ellipsoid, at least $5/6R$ away from the centerline). We define the excursions of a particle, $i$, as:
\begin{align}
E_i(t) \;=\; \max_{0 \le t' \le t} \left| x_i(t') - x_i(0) \right| \\
E_i^{(r)}(t) \;=\; \max_{0 \le t' \le t} \left| r_i(t') - r_i(0) \right|,
\end{align}
where the $x$-axis is the long axis of the ellipsoid and $r(x,y,z)=\sqrt{y^2+z^2}$ measures distance from the $x$-axis.

We found that the traveling wave regime greatly increases both long-axis and radial transport properties  [\cref{fig:figure4b}, A]. For example, in the traveling wave regime (from $\bar{\sigma}$'s $90-110$ at a $\bar{\rho}$ value of $15$), after around $8$ hours of streaming the mean excursion distance along the long-axis is greater than the semi-major length of the ellipsoid. The maximum excursion distance is on the order of the length of the cell, carrying particles from one pole to the opposite pole  [\cref{fig:figure4b}, B]. For context, the duration of cytoplasmic streaming in the oocyte is on the order of $12$ hours. All the regimes with fluid flows transport particles radially, but the waving flows bring cortically deposited particles to the center far more quickly than stable streaming flows. By $6$ hours, the mean radial excursion for a particle is greater than $2R/3$ whereas for stable streaming that value is less than $R/6$. 

While the transport properties continue to increase as a function of $\bar{\sigma}$ even as the traveling wave regime transitions to the beating regime, we found that there is a $\bar{\sigma}$ at which the beating regime flows became detrimental to particle transport [\cref{fig:figure3}, C.iv, \cref{fig:figure4b}, B]. In these very high $\bar{\sigma}$ flows, particles get trapped in local orbits. This peaked feature of transport ability as a function of $\bar{\sigma}$ may explain why the biological system lives in the traveling wave regime  [\cref{fig:figure4b}, A]. When $\bar{\sigma}$ is too high, not only is the cell expending more energy, the flows generated become worse transporters. And in fact, the biological system may actually live closer to the edge of the beating regime. From microscopy, waves are apparent and consistent. However, over longer imaging durations and during the dumping process, we observed that the waves can change direction. There is never any high angle deviation of the microtubules from the overall flow direction (unlike in the beating regime), but the changing direction of the waves is consistent with greater movement of the defects as seen on the edge of waving and beating. 

Because the biological system needs to both transport maternal RNAs across its long axis and redistribute yolk granules post dumping, and because the waving flows have stronger transport properties than the pure streaming flows, we hypothesize that the biological system lives in the traveling wave regime due to its improved transport and mixing capacity.  

\section*{Discussion}
The interactions between motor-loaded microtubules through the surrounding cytoplasmic fluid have been hypothesized to generate cytoplasmic streaming, which is crucial for organizing large cells such as the \textit{Drosophila} oocyte \cite{Corti1774,Wang2008,Palacios2002,Quinlan2016,cooley1994cytoskeletal}. Our experiments showed that in the \textit{Drosophila} oocyte, microtubule beds are not static, but support traveling waves of deformation. Our statistical analysis of imaging data revealed that the observed dynamics is low-dimensional and can be adequately captured by 3 modes, the first an axisymmetric mode associated with the underlying twister flow, and the other two co-equal contributions from the traveling wave. It is worth noting that these waves, with wavelengths ${\sim}50 \mu \textrm{m}$, are a collective feature of the microtubule bed, and not of the individual microtubules whose lengths are hypothesized to be $20 \mu \textrm{m}$ or less \cite{Dutta2024SelfOrganizedTwisters}. These traveling waves are unlike the metachronal waves of ciliary beds, which are also a collective effect. A single cilium produces no external flow, except through its motion, while the moving cargoes on a microtubule will produce flow whether the microtubule moves or not \cite{Poon2025,Chakrabarti2022}. Furthermore, we demonstrated that traveling waves naturally emerge in a region between the stable streaming and beating regimes of the Twister model. Thus, the observed low-dimensional dynamics are a robust feature of the experimental system and the theoretical model. 

Having found this new regime of microtubule bed dynamics, we sought to understand its biological relevance. While streaming has been known to play a crucial role in the transport of maternally derived molecules such as the mRNA \textit{nos} \cite{Quinlan2016,Forrest2003NanosLocalization}, we showed here that this transport may rest upon flow unsteadiness, i.e. traveling waves. The relation of flow and yolk suspension is much less studied. Our experiments characterizing the distribution of yolk granules, with and without streaming, showed that cytoplasmic streaming is necessary for the uniform distribution of yolk by the end of oogenesis. These experiments suggest, in particular, that fast nurse cell dumping pushes yolk to the periphery in the absence of streaming. While the time-scale for dumping is on that of the twister circulation time -- suggesting that incoming yolk-laden cytoplasm could be mixed rapidly into fresh cytoplasm by streaming -- we also showed that traveling waves induce inward radial flows that would also promote yolk resuspension. In summary, we hypothesize that the functional role of traveling waves is to enhance (or enable at the relevant timescales) transport of both yolk granules and mRNAs.

Our current observations leave open other questions about the traveling wave regime and its effects in a crowded cytoplasm. Future work is needed to understand what controls the size of the parametric regime of traveling waves and how the biological system ensures that it operates in that traveling wave regime. Additionally, there is work to be done to characterize both the bifurcation from stable streaming to traveling waves and the breakdown of waves into beating. Lastly, there are open questions about the mixing properties of cytoplasmic streaming. Our simulational model, which captured complex fluid-microtubule interactions, considered the enhanced transport properties of zero-volume tracer particles. This is reasonable for small mRNA molecules, but yolk granules are several $\mu\textrm{m}$ in size. We are currently building more sophisticated theoretical and computational models to investigate the effects of a dense yolk plasm on cytoplasmic rheology and transport during streaming. 

\section{Materials and Methods}
\subsection{Jupiter-GFP Live Imaging}
Jupiter-GFP (ZCl2183) flies were anesthetized on CO$_2$ pads and their ovaries were dissected in Schneider's Mix media supplemented with Insulin, FBS, and streptomycin/penicilin, and pH adjusted as described by \cite{prasad2007protocol}. The egg chambers were imaged in MatTek 35mm glass bottom culture dishes 
using a Nikon AX laser scanning confocal microscope and the NIS-Elements software at the CCB Scope Observatory at the Flatiron Institute. Imaging was performed using a 40x/1.2 silicon oil objective lens. Pinhole settings ranged from 1.0 to 1.2 Airy units. Excitation of the fluorophore, GFP, was performed at 489 nm. 3D volumes 10-15 $\mu \mathrm{m}$ deep were acquired at 1 $\mu \mathrm{m}$/step. 

\subsection{Image Analysis and Modal Decomposition}
\label{image_analysis_and_modal_decomposition}
A max projection of $5-10 \mu \mathrm{m}$ from the aligned microtubule bed was used to generate a single-plane movie of microtubule waving. A bank of Gabor filters, each composed of a sinusoid multiplied by a Gaussian, were rotated $180^\circ$ degrees in $1^\circ$ increments on a grid of size $20-40$ pixels depending on the image. The response with the highest resulting magnitude was used to extract the orientation vector per grid point. Magnitudes below a threshold were excluded as noise. 

SVD analysis was used to identify the dominant spatiotemporal modes in the extracted orientations. A matrix $A$ is formed with each column taking the orientational data at a given time $t$. To form each column, the orientation vectors are normalized and the flattened $v_{x}$ and $v_{y}$ components at each gridpoint are stacked. $A$ was decomposed into $U\Sigma V^T$ using the singular value decomposition. The resulting spatial matrix, $U$, was unfolded onto the grid to visualize each spatial mode.

While the flow direction of the movies was largely perpendicular to the long axis of the cell (axisymmetric flow), there were movies in which a vortex or defect was visible. Such movies were excluded from the analysis. Additionally, the oocyte grew during the course of the movies. To remove effects of boundary change and volume increase, movies were analyzed for the duration during which the oocyte volume did not substantially change (anywhere between $3-10$ minutes) so that the grid size could stay consistent for the duration of the movies. Select movies were analyzed for both their initial and final 5 minutes to check whether the results of SVD were consistent. 

\subsection{Cryosection and Yolk Granule Imaging}
Oregon-R (OR) WT flies were anesthetized on CO$_2$ pads and their ovaries were dissected in Schneider's Mix media supplemented with Insulin, FBS, and streptomycin/penicilin, and pH adjusted as described by \cite{prasad2007protocol}. To visualize the consequences of no streaming, stage 10B oocytes were selected and incubated for 8 hours with gentle rotation and colchicine treatment as described in \cite{Forrest2003NanosLocalization}. Additionally stage 12-14 oocytes were selected both conditions were fixed for 20 minutes with 38\% formaldehyde in PBS. 

For cryosectioning, the oocytes were dehydrated with a 30\% sucrose solution overnight and frozen/embedded in Optimal Cutting Temperature media. $7 \ \mu \textrm{m}$ sections were cut using a Leica CM3050S Cryostat in the Princeton University Histology Core. The slices were placed on glass slides and imaged using a Nikon AX laser scanning confocal microscope and the NIS-Elements software at the CCB Scope Observatory at the Flatiron Institute. Imaging was performed using a 40x/1.2 silicon oil objective lens. The smallest pinhole size of $6 \ \mu \textrm{m}$ was used. Excitation of the autofluorescent yolk was performed at 405 nm. 3D volumes 10-15 $\mu \mathrm{m}$ deep were acquired at 0.35 $\mu \mathrm{m}$/steps. 

\subsection{Numerical Simulations}
\label{methods:numerics}
The oocytal dynamics simulations in this dissertation use \emph{SkellySim2.0}, a codebase for simulating large numbers of semi-flexible and inextensible filaments hydrodynamically interacting through a Newtonian fluid (at zero Reynolds number) in which they are immersed. This code, currently unreleased as it is still in early and active development, is the successor to the publicly available \emph{SkellySim} \cite{skellysim}, which provided the source for previous related investigations \cite{Dutta2024SelfOrganizedTwisters}. As in SkellySim, the microtubules are treated as slender and inextensible Euler elastica, and hydrodynamic interactions are treated approximately using slender body theory \cite{Tornberg2004}. However, SkellySim2.0 has several substantial improvements over SkellySim which enabled new parameter regimes to be explored over longer timescales, revealing new phenomena (waves) and enabling long-time analysis of mixing and transport. These improvements include:
\begin{enumerate}
\item The stiff fourth-order equations of motion are discretized using a well-conditioned integral reformulation, with a spectral representation based on Chebyshev polynomials \cite{vicentehurryitup};
\item Steric interactions are included using stiff repulsive potentials, discretized implicitly in time to permit relatively large timesteps;
\item The near-singular integrals that arise due to microtubules being close to each other or to the cortex are computed accurately using Gauss-Kronrod based adaptive interpolatory quadratures (for microtubules), or through the internal adaptive routines in Inti.jl \cite{anderson2024inti}, (for the cortex);
\item Support for high-order adaptive timestepping (for all simulations in this manuscript, the classic third-order backward differentiation formula was used);
\item The resulting large and dense nonlinear systems of equations are not linearized, but instead inverted using a preconditioned exact Newton-Krylov scheme, with Jacobians and Jacobian-vector products computed via automatic differentiation, supported by the ForwardDiff.jl package and related ecosystem within Julia \cite{RevelsLubinPapamarkou2016}.
\end{enumerate}
Collectively, these improvements allow faster simulations in the higher density ($\rho)$ and higher forcing ($\sigma$) regimes where the dynamics become unsteady enough to drive rapid transport and mixing. The simulations were run across a range of $\rho$ and $\sigma$ values to generate phase diagrams for a sphere and two different aspect ratio ellipsoids.

To characterize the traveling wave, any microtubule whose base was located between $-1<x<1$ was selected. The semi-major axis of the ellipsoid varied from $3-7$.  Each microtubules position is stored at 36 points along its arc-length (specifically, the first-kind Chebyshev nodes). The microtubules have slenderness ratio, $r/L=0.001$. Using estimates of microtubule length and flexural rigidity, the simulations are non-dimensionaled by $\tau_r=12,642 \ \textrm{s}$ \cite{Gittes1993,Lu2022}. To calculate the angle of the microtubule in the simulation, the vector $(\mathbf{X}[36]-\mathbf{X}[15])$ was used. $\mathbf{X}[15]$ was used because it was the point at which the average microtubule became bent. Because flow in an ellipsoid is axisymmetric and the long axis of the ellipsoid was set as the $x$-axis, the angle between $(\mathbf{X}[36]-\mathbf{X}[15])$ and $(1,0,0)$ was calculated as $\phi$. The $\phi$ value was averaged around the cross sectional circle $(y/b)^2+(z/b)^2=1$ at the mid-plane of the ellipsoid where $b$ was the semi-major axis. $\phi$ as a function of $\theta$ (where $\theta$ parameterized the circle) was calculated for every time point to generate the kymographs. Traveling wave speeds were calculated using the peak/trough speed from each $\phi$ vs. $\theta$. 

Spatial modes were extracted from the numerical simulation data in a manner analogous to the experimental data. Microtubules below some $z$-plane (from 1/6 to 1/3 the oocyte's height) were projected onto the $xy$-plane as would be visualized by a cortical slice of the oocyte on a microscope. The angle given by the microtubule projections was interpolated onto a structured grid for each time point, SVD was performed [see Methods]. The spatial modes identified from the numerical data in this way are compared to those from the experimental data in [see S.I.].

\section{Acknowledgments}
We wish to thank Brato Chakrabarti, Wen Lu, and Vladimir Gelfand for valuable discussions, and Chris Edelmaier and Robert Blackwell for help with code development. We acknowledge support from the CCB${}_\textnormal{X}$ program of the Center for Computational Biology of the Flatiron Institute. The experimental observations reported in this work were performed in the CCBScope Observatory at the Flatiron Institute. This material is based upon work supported by the National Science Foundation Graduate Research Fellowship under Grant No. DGE-2039656 (A.O.J.) and by the National Institutes of Health under Award Number R35-GM067758. (E.R.G.). The Flatiron Institute is a division of the Simons Foundation.

\bibliographystyle{plain}
\bibliography{references}


\clearpage

\begin{center}
    {\LARGE\bfseries Supporting Information\par}
    \vspace{1em}

    {\Large\bfseries
    Traveling Waves Enhance Transport during Cytoplasmic Streaming in
    \textit{Drosophila} Oogenesis\par}

    \vspace{1em}

    {\large
    Olenka Jain, David B. Stein, Reza Farhadifar, Elizabeth R. Gavis,\\
    Stanislav Y. Shvartsman, Michael J. Shelley\par}

    \vspace{0.75em}

    {\small
    Correspondence: Michael J. Shelley\\
    \texttt{mshelley@flatironinstitute.org}\par}
\end{center}

\vspace{1cm}

\begin{figure}[p]
\centering
\includegraphics[width=\textwidth]{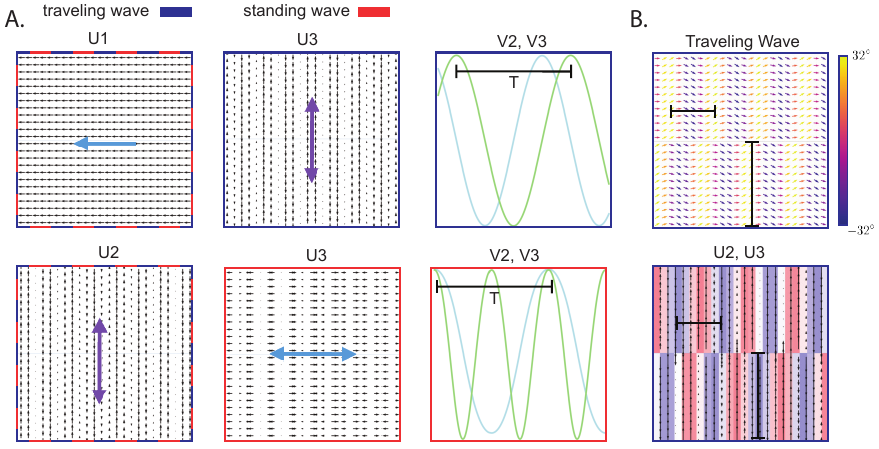}
\caption{A. Singular value decomposition performed on representative traveling and standing waves. For both a low amplitude/high wavelength traveling wave and a low amplitude/high wavelength standing wave, the first spatial mode, $U_1$, is a vector in the direction of travel, $\vec{\mathbf{v}}$. For both wave types, $U_2$ alternates between patches of $\vec{\mathbf{v}}+\frac{\pi}{2}$ and $\vec{\mathbf{v}}-\frac{\pi}{2}$. The difference between wave types appears in $U_3$. For the traveling wave, $U_3$ is a phase shift of $U_2$; for the standing wave, $U_3$ has vectors parallel or anti-parallel to $\vec{\mathbf{v}}$. The time dependent amplitudes likewise reveal the difference between traveling and standing waves. For the traveling wave, $V_2$ and $V_3$ have equal period and are $\pi/2$ out of phase. For a standing wave, $V_3$ is double the period of $V_2$. For both types of waves, the period is given by $V_2$; for a traveling wave this can be derived analytically using a small angle approximation. B. For a traveling wave, the size of the $\vec{v}+\frac{\pi}{2}$ and $\vec{v}-\frac{\pi}{2}$ regions gives the wavelength of the underlying traveling wave, and the length of that patch is set by the phase.}
\end{figure}

\begin{figure}[p]
\centering
\includegraphics[width=\textwidth]{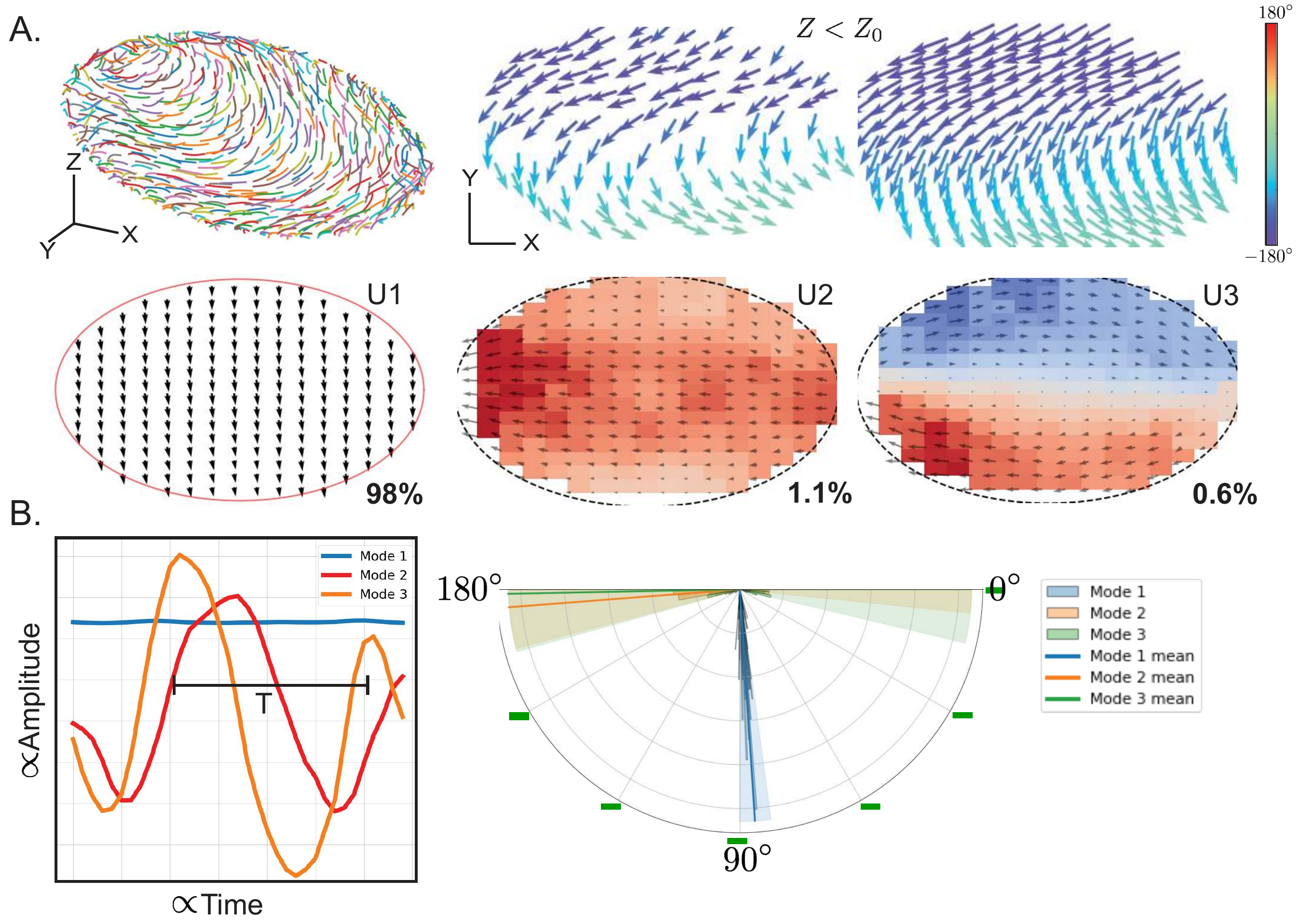}
\caption{A. Singular Value Decomposition performed on a simulated microscopy slice. Fibers from the 3D simulation below some $Z$ height were projected onto the $X-Y$ plane where the $X-$axis was the long-axis of the ellipsoid. The fiber angle was interpolated onto a structured grid and used as input for the SVD. The results showed consistency with the experiments both in the structure of the spatial modes. The first mode explained over $90\%$ of the variance and contained vectors uniformly in the direction of solid body rotation (up to a sign change). The next two spatial modes were about equal in their importance, with vectors perpendicular to the first mode and oscillating time dependent amplitudes which were $\frac{\pi}{2}$ out of phase, but had the same period. The analysis of spatial structure in the top three modes is summarized by the circular histogram.}
\end{figure}

\begin{figure}[p]
\centering
\includegraphics[width=\textwidth]{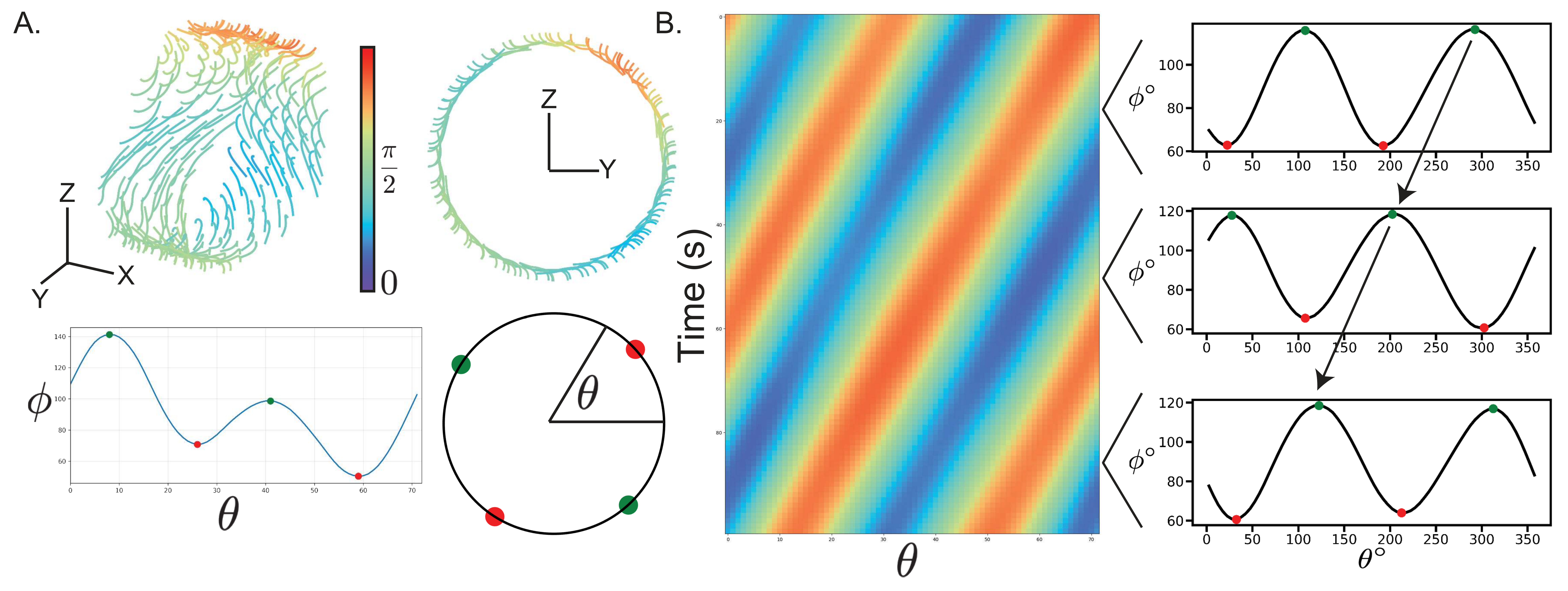}
\caption{A. The center belt of fibers from $-1<X<1$ was used to characterize the traveling wave in all ellipsoids, whose long axis was set as the $X-$axis. The average angle of each fiber relative to the long axis, $\phi$, was plotted as a function of $\theta$ of the cross-sectional circle. The peaks and troughs are shown in red and green. B. A kymograph of was generated where the maximum and minimum of the color bar are set by the maximum and minimum angle deviation from $\frac{\pi}{2}$.}
\end{figure}

\begin{figure}[p]
\centering
\includegraphics[width=\textwidth]{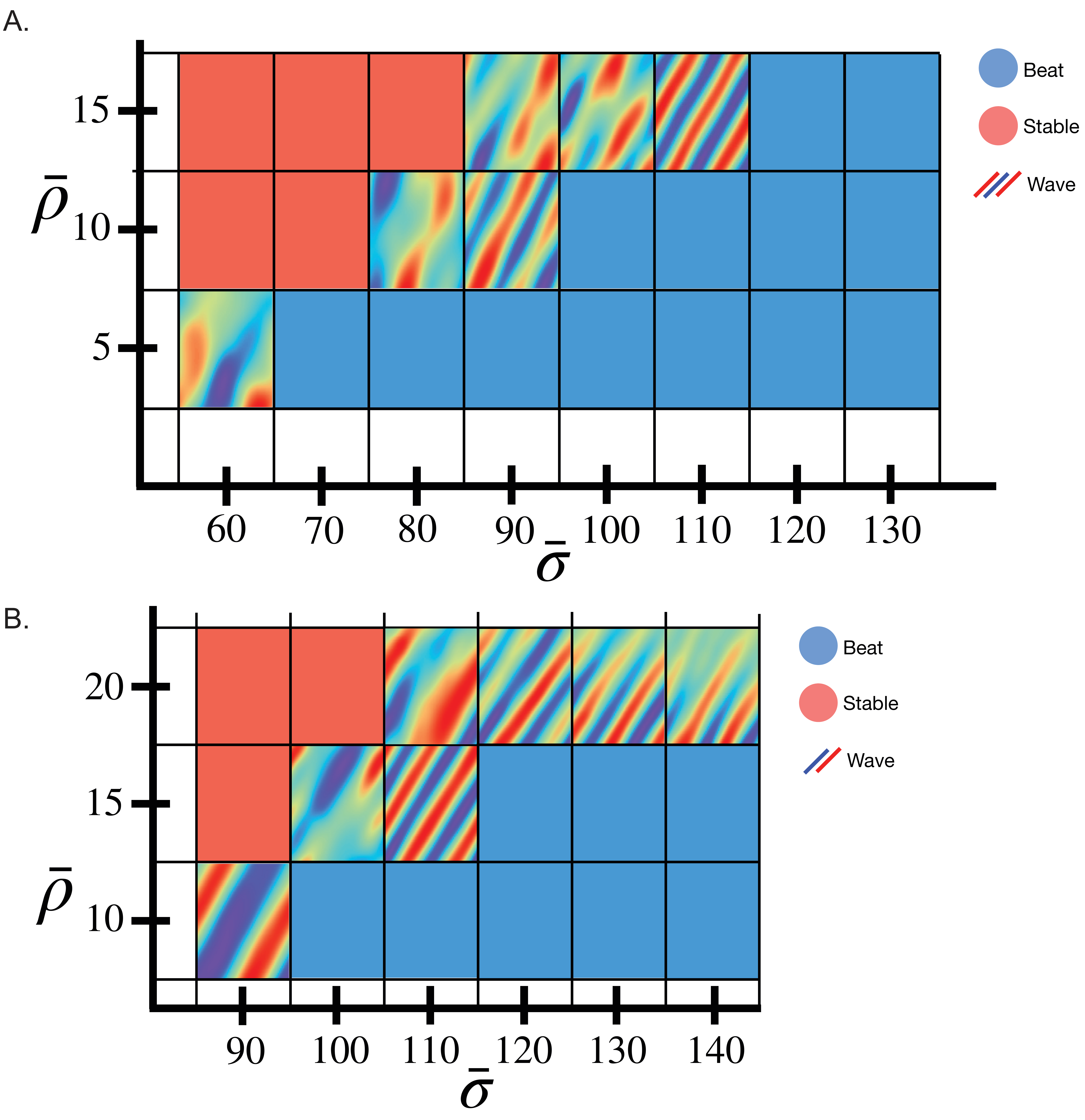}
\caption{A. Simulations were run across a range of $\bar{\rho}$ and $\bar{\sigma}$ parameters for an ellipsoid with semi-major and semi-minor axes: $A=5,B=3,C=3$. Waves are illustrated by kymographs. B. Simulations were run across a range of $\bar{\rho}$ and $\bar{\sigma}$ parameters for an ellipsoid with semi-major and semi-minor axes: $A=7,B=3,C=3$. For both ellipsoids, waving dynamics were found between the stable and beating regimes.}
\end{figure}

\clearpage

\section{Supporting Movies}

\movie{Waving dynamics in a Stage 10B oocyte (1) visualized with Jupiter-GFP line.}

\movie{Waving dynamics in a Stage 10B oocyte (2) visualized with Jupiter-GFP line.}

\movie{Waving dynamics in a Stage 12 oocyte visualized with Jupiter-GFP line.}

\movie{Waving dynamics in a Stage 10B oocyte about to dump where the defect has moved from the poles and the waves appear more consistent with waves on the edge of streaming and beating.}

\movie{Radial transport in simulations with $\bar{\rho}=15$ and $\bar{\sigma}=50$, in the steady streaming regime, with a bi-toroidal component pushing particles near the pole inward along the axis. Tracer particles are initialized randomly using rejection sampling, with $-1\leq x\leq 1$, $r>2.5$, and within the ellipsoid (i.e. near the center along the axis and close to the periphery). Their motion is visualized over one hour.}

\movie{Radial transport in simulations with $\bar{\rho}=15$ and $\bar{\sigma}=80$, in the steady streaming regime, with a bi-toroidal component pushing particles near the pole outward along the periphery. Tracer particles are initialized randomly using rejection sampling, with $-1\leq x\leq 1$, $r>2.5$, and within the ellipsoid (i.e. near the center along the axis and close to the periphery). Their motion is visualized over one hour.}

\movie{Radial transport in simulations with $\bar{\rho}=15$ and $\bar{\sigma}=120$, in the waving regime. Tracer particles are initialized randomly using rejection sampling, with $-1\leq x\leq 1$, $r>2.5$, and within the ellipsoid (i.e. near the center along the axis and close to the periphery). Their motion is visualized over one hour.}

\movie{Radial transport in simulations with $\bar{\rho}=15$ and $\bar{\sigma}=200$, in the beating regime. Tracer particles are initialized randomly using rejection sampling, with $-1\leq x\leq 1$, $r>2.5$, and within the ellipsoid (i.e. near the center along the axis and close to the periphery). Their motion is visualized over one hour.}

\movie{Axial transport in simulations with $\bar{\rho}=15$ and $\bar{\sigma}=50$, in the steady streaming regime, with a bi-toroidal component pushing particles near the pole inward along the axis. Tracer particles are initialized randomly using rejection sampling, with $x > -4$ and within the ellipsoid (i.e. near the left cap). Their motion is visualized over one hour.}

\movie{Axial transport in simulations with $\bar{\rho}=15$ and $\bar{\sigma}=80$, in the steady streaming regime, with a bi-toroidal component pushing particles near the pole outward along the periphery. Tracer particles are initialized randomly using rejection sampling, with $x > -4$ and within the ellipsoid (i.e. near the left cap). Their motion is visualized over one hour.}

\movie{Axial transport in simulations with $\bar{\rho}=15$ and $\bar{\sigma}=120$, in the waving regime. Tracer particles are initialized randomly using rejection sampling, with $x > -4$ and within the ellipsoid (i.e. near the left cap). Their motion is visualized over one hour.}

\movie{Axial transport in simulations with $\bar{\rho}=15$ and $\bar{\sigma}=120$, in the beating regime. Tracer particles are initialized randomly using rejection sampling, with $x > -4$ and within the ellipsoid (i.e. near the left cap). Their motion is visualized over one hour.}

\end{document}